\documentclass{revtex4}
\def\giorno{31/10/2017}

\usepackage{color}
\usepackage{amsmath,amsfonts,amssymb}

\def\b{\beta}

\def\de{\delta}   
\def\eps{\varepsilon}
\def\vphi{\varphi}

\def\s{\sigma}

\def\om{\omega}

\def\vphi{\varphi}

\def\L{\mathcal{L}}

\def\pa{\partial}

\def\d{{\rm d}}       
\def\w{\wedge}

\def\o+{\oplus}

\def\ss{\subset}

\def\<{\langle}
\def\>{\rangle}

\def\({\left(}
\def\){\right)}
\def\[{\left[}
\def\]{\right]}
\def\=#1{\bar #1}
\def\~#1{\widetilde #1}
\def\wt#1{\widetilde #1}
\def\.#1{\dot #1}
\def\^#1{\widehat #1}
\def\wh#1{\widehat #1}
\def\"#1{\ddot #1}

\def\interno{\hskip 2pt \vbox{\hbox{\vbox to .18
truecm{\vfill\hbox to .25 truecm
{\hfill\hfill}\vfill}\vrule}\hrule}\hskip 2 pt}

\def\eeq{\end{equation}}
\def\beq{\begin{equation}}

\def\beql#1{\begin{equation} \label{#1}}

\def\eqref#1{(\ref{#1})}

\def\EOR{ \hfill $\odot$ \medskip}

\def\EOP{ \hfill $\triangle$ \medskip}

\begin{document}

\title{On Lie-point symmetries for Ito stochastic differential equations}

\author{G. Gaeta\thanks{ORCID 0000-0003-3310-3455; e-mail: {\tt giuseppe.gaeta@unimi.it}} \ \ and C. Lunini\thanks{ORCID 0000-0002-2643-1226; e-mail: {\tt claudia.lunini@studenti.unimi.it}}}
\affiliation{Dipartimento di Matematica, Universit\`a degli Studi di
Milano, via Saldini 50, 20133 Milano (Italy)} 

\date{\giorno }

\begin{abstract}

In the deterministic realm, both differential equations and symmetry generators are geometrical objects, and behave properly under changes of coordinates; actually this property is essential to make symmetry analysis independent of the choice of coordinates and applicable. When trying to extend symmetry analysis to stochastic (Ito) differential equations, we are faced with a problem inherent to their very nature: they are not geometrical object, and they behave in their own  way (synthesized by the Ito formula) under changes of coordinates. Thus it is not obvious that symmetries are preserved under a change of coordinates. We will study when this is the case, and when it is not; the conclusion is that this is always the case for so called \emph{simple} symmetries. We will also note that Kozlov theory relating symmetry and integrability for stochastic differential equations is confirmed by our considerations and results, as symmetries of the type relevant in it are indeed of the type preserved under coordinate changes.

\end{abstract}

\maketitle

\section{Introduction}

Symmetry methods are our most powerful tool in studying nonlinear deterministic differential equations \cite{AVL,CG,Olv1,Olv2,Ste}. It is thus natural to attempt using them also for the study of \emph{stochastic} differential equations (SDEs).

In fact, there is by now a substantial literature concerned with symmetries of SDEs; the reader is referred to the extensive list of references in the recent review \cite{GGPR} for a (more or less complete) bibliography.

It should be mentioned that by a SDE (with no further specification) we will always intend an Ito stochastic equation\footnote{More precisely, a forward one; a similar (equivalent) theory could be built for backward Ito equations.}; as well known, these are the natural ones to be considered in view of probabilistic properties \cite{Eva,Fre,IW,Kam,Oks,Stroock}, but other types of SDEs are also considered in the literature. In particular, physicists also like to consider Stratonovich SDEs. These, at difference with Ito ones, $(i)$ are symmetric under time reversal, and $(ii)$ transform under changes of coordinates in the ``usual'' way, i.e. according to the chain rule -- and we will see in the following they also enjoy other favourable properties from the point of view of symmetry. On the other hand, they present several problems from the point of view of probabilistic foundations \cite{Eva,Fre,IW,Kam,Oks,Stroock}, so that the mathematical literature prefers to consider Ito equations. It is also well known that for each Ito equation there is an ``equivalent'' Stratonovich equation \cite{Eva,Fre,IW,Kam,Oks}. The exact nature of this equivalence is however rather delicate when one looks carefully at it; see e.g. the discussion in \cite{Stroock}. A physicist's point of view focusing on sample paths and Feynman's path integrals is given in \cite{Schu}.

The study of symmetry of SDEs concentrated first on Stratonovich equations\footnote{We ignore the reasons for this, albeit it is possible it is due to the fact symmetries are primarily of interest to physicists, and as mentioned above there are good physical reasons to consider Stratonovich equations.}, but then extended to Ito equations as well, and actually mainly focused on these.

We also stress that we are always considering the stochastic version of an ODE; that is, we are not considering stochastic PDEs. Thus when thinking of the (much better known and understood) deterministic counterpart, we should think of the theory applying for ODEs.

Quite surprisingly, it appears that studies have so far not tackled an essential problem, concerned with an equally essential difference between stochastic (Ito) equations and deterministic ones.

In fact, it is well known -- and surely well known to the readers of this special issue of JNMP, devoted to symmetry of nonlinear equations -- that a deterministic differential equation $E$ has a geometrical meaning as a submanifold $S_E$ in a certain jet bundle; similarly, the generators of Lie-point symmetries are vector fields $X$ in the base manifold for the same bundle, extended to vector fields $\^X$ in the total space for the bundle by the \emph{prolongation} operation. Symmetry corresponds to these prolonged vector fields being tangent to the manifold representing the equation \cite{AVL,CG,Olv1,Olv2,Ste}.

Such a relation is a geometric one, and hence it is of course \emph{independent} of the coordinates we are using. This is essential in several aspects; focusing on ODEs, in particular, we know that symmetry reduction under a vector field $X$ is based on using symmetry adapted coordinates -- in which the symmetry property guarantees the reduction is immediate (and reconstruction amounts to a quadrature); see e.g. \cite{AVL,CG,Olv1,Olv2,Ste} for details. In case there are several Lie-point symmetries, we are automatically guaranteed these will still be present in the new coordinates\footnote{As well known, the preservation of symmetries is instead not automatic under the reduction operation; this depends on the Lie algebraic structure of the symmetry algebra, and on performing the symmetry reduction ``in the right order'' \cite{AVL,CG,Olv1,Olv2,Ste}.}, just due -- as mentioned a few lines above -- to the fact vector fields, submanifolds, and hence tangency relations among these, are geometrical objects, independent of the coordinates representation.

But when we enter in the stochastic realm, we get immediately a serious difference. Lie-point symmetry generators are still vector fields, and surely they behave ``properly'' (i.e. according to the familiar chain rule) under changes of coordinates. Stratonovich differential equations also behave properly (in the same sense) under changes of coordinates, and actually this was the reason which led Stratonovich to formulate his theory. But Ito equations do \emph{not} transform according to the chain rule; under a change of coordinate they transform according to the Ito rule. In other words, Ito equations are \emph{not covariant}.\footnote{This problem
seems not to have been discussed in the literature (including in
contributions by the present authors), possibly because it is
``intuitively clear'' that symmetries are such independently of
the coordinates; as we will see in a moment, such an intuition is
indeed correct -- at least for simple symmetries.}

Thus, we expect some trouble can arise when we consider symmetry properties \emph{and} change of coordinates for an Ito equation (or systems thereof). This is not just a mathematical curiosity, but an essential point when we try to apply symmetry theory to SDEs. A substantial (and for us, motivating) example is given by Kozlov's theory \cite{Koz1,Koz2,Koz3}, establishing the connection between integrability of a SDE and its symmetry properties, and more in general the possibility to reduce a system of SDEs making use of its simple (he exact meaning of this term will be explained below) symmetries; this is based on changes of coordinates, exactly as in the deterministic case. (We anticipate that Kozlov's theory is perfectly fine, thanks to some special property of the type of transformations which are relevant in it.)

The purpose of this note is to clarify the point raised above, i.e. study how the symmetry properties of a SDE change under changes of coordinates.We have a special interest, of course, in the case where these do \emph{not} change with a change of coordinates -- after all, this is what makes the theory really applicable -- and thus we will devote special attention to this question. The result is that, luckily, a large set of symmetries of an Ito equation are indeed preserved under changes of coordinates; not all of them, but all the ``useful'' ones.

We would like to stress that we will only consider ``proper'', i.e. first order, (systems of) Ito equations. The literature also considers (systems of) higher order SDEs and their symmetries \cite{GGPR,SMS}; we will not consider these, neither other possible generalizations of the setting we have chosen to study.

\section{Symmetry of stochastic differential equations}
\label{sec:symmsde}

In the case of deterministic differential equations\footnote{By
``differential equation'' we will always mean possible a vector
one, i.e. a system of differential equations, unless the contrary
is explicitly stated.} $\Delta$ of order $n$ in the basis manifold
$M$ (this include dependent and independent variables) there is a
well developed theory identifying them with suitable submanifolds $S_\Delta$
in the Jet manifold $J^n M$; a Lie-point symmetry is then a vector
field $X$ in $M$ which, when \emph{prolonged} to $J^n M$, is
tangent to $S_\Delta \ss J^n M$.

In the case of a (Ito) stochastic differential equation (SDE)
\beql{eq:Ito} d x^i \ = \ f^i (x,t) \, d t \ + \ \s^i_{\ j} (x,t)
\, d w^j \ , \eeq such a \emph{geometrical} interpretation of the
equation is missing\footnote{It is instead possible in the case of
Stratonovich equations, see Section \ref{sec:redstrat} below.},
and we have to resort to a purely \emph{algebraic} notion of symmetry.

It should be mentioned that one could, in principles, consider
general mappings in the $(x,t,w)$ space -- albeit with some
restrictions if these have to make physical sense, see e.g.
\cite{GGPR,GS} -- but in the context we are presently interested in,
the only relevant ones are \emph{simple} maps. By this we mean maps leaving the time variable $t$ and the Wiener processes $w$ unchanged. This simplifies considerably our task.

\medskip\noindent
{\bf Remark 1.} The reason to focus on simple maps is their role in application. In fact, Kozlov \cite{Koz1,Koz2,Koz3} showed that the symmetries which are useful to integrate a system of SDEs are \emph{only} the simple ones. We refer the reader to his papers, or to the review \cite{GGPR}, for details on this theory. \EOR
\bigskip

In \eqref{eq:Ito} and below, $t \in R$, $x \in R^n$ (more general
situations, with $x$ taking values in a smooth manifold, are also
possible -- but again we will not consider these here), $f$ and $\s$ are
smooth vector and matrix functions of their arguments, and $w^j$
($j = 1,...,m$) are independent standard Wiener processes.

\subsection{Simple deterministic symmetries}

Let us first consider a smooth\footnote{By ``smooth'' we
will always mean $C^\infty$, albeit in several steps it would be
sufficient to consider $C^2$ smoothness. As already stated, by
``simple'' we mean a map which acts only on the spatial variables,
leaving the time $t$ (and the Wiener processes) unchanged. We will
refer to such a map as ``deterministic'' (as opposed to the
``random'' ones to be considered below, in which the Wiener
processes $w$ play a role) in that it only involves the $(x,t)$
variables, albeit of course when we introduce an Ito equation its solution
$x(t)$ is a stochastic process.} map \beql{eq:smap} (x,t) \ \to
(\wt{x} , t ) \ , \ \ \wt{x} = \wt{x} (x,t) \ ; \eeq such maps will be denoted as \emph{simple}. This map \eqref{eq:smap} induces
a map $d x \to d \wt{x}$ and hence a map on SDEs \eqref{eq:Ito};
in particular if \eqref{eq:smap} is (locally) inverted to give
\beq x \ = \ \Phi (\wt{x},t ) \ , \eeq then by Ito formula \beq d
x^i \ = \ \( \frac{\pa \Phi^i}{\pa \wt{x}^j} \) \, d \wt{x}^j \ +
\ \( \frac{\pa \Phi^i}{\pa t} \) \, d t \ + \ \frac12 \( \frac{\pa^2
\Phi^i}{\pa \wt{x}^j \pa \wt{x}^k} \) \wt{\s}^j_{\ \ell}
\wt{\s}^k_{\ \ell} \, d t \ ; \eeq similarly the functions $f^i
(x,t)$ and $\s^i_{\ j} (x,t)$ are mapped into functions $\wt{f}^i
(\wt{x},t ) $ and $\wt{\s}^i_{\ j} (\wt{x} , t )$. In this way,
\eqref{eq:Ito} is mapped into a new Ito equation
\beql{eq:Ito2} d \wt{x}^i \ = \ \wh{f}^i
(\wt{x},t) \, d t \ + \ \wh{\s}^i_{\ j} (\wt{x},t) \, d w^j \ ; \eeq
note that here the $\wh{f}$ and $\wh{\s}$ take into account not
only the change of their variables, but also the contribution
arising from $d x$ expressed in the new variables via the Ito
formula.

We say that \eqref{eq:smap} is a symmetry for \eqref{eq:Ito} if
\eqref{eq:Ito2} is identical to \eqref{eq:Ito}, i.e. if the ($n +
m \cdot n)$ conditions \beq \wh{f}^i (x,t) \ = \ f^i (x,t) \ , \ \
\wh{\s}^i_{\ j} (x,t) \ = \ \s^i_{\ j} (x,t) \eeq are satisfied
identically in $(x,t)$.

We are specially interested in the case where the map
\eqref{eq:smap} is a near-identity one, $\wt{x}^i = x^i + \eps
(\de x)^i$ and can hence be seen as the infinitesimal action of a
vector field $X$ on $M$, \beql{eq:X} X \ = \ \vphi^i (x,t) \,
\frac{\pa}{\pa x^i} \ \equiv \ \vphi^i (x,t) \, \pa_i \ . \eeq
(Note that $X$ has no component along the $t$ variable; this
corresponds to the restriction made above, see eq.\eqref{eq:smap},
for simple symmetries.) In this case we speak of a
\emph{Lie-point} (simple) symmetry.

Proceeding in this way (the reader is referred to \cite{GRQ2,GGPR,GRQ1} for
details and explicit computations) we obtain the \emph{determining
equations for (simple, deterministic) Lie-point symmetries of
SDEs}; these read
\begin{eqnarray}
\vphi^i_t \ + \ f^j \, (\pa_j \vphi^i) \ - \ \vphi^j \, (\pa_j f^i) &=&
- \, \frac12 \ (\Delta \vphi^i ) \ , \label{eq:deteqIto1}  \\
\s^j_{\ k} \, (\pa_j \vphi^i) \ - \ \vphi^j \, (\pa_j \s^i_{\ k} )
&=& 0 \ . \label{eq:deteqIto2} \end{eqnarray} Here and below we
denote by $\Delta$ the \emph{Ito Laplacian}, which is in general (for functions possibly depending also on the $w^k$, see next subsection) defined
as \beql{eq:Delta} \Delta f \ = \ \delta_{ik} \
\[ \( \frac{\pa^2 f}{\pa w^i \pa w^k} \) \ + \ \s^i_{\ j} \s^k_{\
m} \, \( \frac{\pa^2 f}{\pa x^j \pa x^m} \) \ + \ \s^i_{\ m}  \( \frac{\pa^2 f}{\pa x^i \pa w^m} \) \ + \ \s^i_{\ j}  \( \frac{\pa^2 f}{\pa x^j \pa w^k} \) \] \ . \eeq

\medskip\noindent
{\bf Remark 2.} In previous works of ours \cite{GGPR,GS} the mixed derivative term was missing in the definition of $\Delta$; this entails that the concrete computations given there have to be revised, but also led to wrong conclusions concerning the relation between symmetries of Ito versus Stratonovich equations; see Section \ref{sec:ItoStrat} (and Remark 3) in this respect. \EOR

\subsection{Simple random symmetries}

We can consider more general transformations, involving also the $w^k$ variables; these will be called \emph{random maps} \cite{Arnbook,ArnImk,GS}. In particular, for the sake of our investigation we can consider smooth \emph{simple random maps},
\beql{eq:srmap} \( x,t;w \) \ \to \( \wt{x}
, t ; w \) \ , \ \ \wt{x} = \wt{x} (x,t;w) \ ; \eeq
and the corresponding symmetries, i.e. \emph{simple random symmetries}.
These are the vector fields \beql{eq:Xran} X \ = \ \vphi^i (x,t;w)
\, \pa_i \eeq leaving the equation \eqref{eq:Ito} invariant

It is shown in \cite{GS} that the \emph{determining equations for
simple random Lie-point symmetries of an Ito SDE} \eqref{eq:Ito} read
as
\begin{eqnarray}
(\pa_t \vphi^i) \ + \ f^j \, (\pa_j \vphi^i) \ - \ \vphi^j \, (\pa_j f^i) &=&
- \, \frac12 \ (\Delta \vphi^i ) \ , \label{eq:deteqItoR1} \\
(\^\pa_k \vphi^i) \ + \ \s^j_{\ k} \, (\pa_j \vphi^i) \ - \
\vphi^j \, (\pa_j \s^i_{\ k} ) &=& 0 \ ; \label{eq:deteqItoR2}
\end{eqnarray}
here we have used the shorthand notation
\beq \^\pa_k \ := \ \frac{\pa}{\pa w^k} \ . \eeq

\section{Symmetry of Ito versus Stratonovich  equations}
\label{sec:ItoStrat}

It is well known \cite{Eva,Fre,IW,Kam,Oks,Stroock} that to each Ito equation \eqref{eq:Ito} corresponds a \emph{Stratonovich stochastic differential equation}
\beql{eq:strato} d x^i \ = \ b^i (x,t) \, d t \ + \ \s^i_{\ k}
(x,t) \circ d w^k \ , \eeq where the functions $f^i$ appearing in
\eqref{eq:Ito} and the functions $b^i (x,t)$ appearing here are
related by \beql{eq:itostrat} f^i (x,t) \ = \ b^i (x,t) \ + \ \frac12 \ \[ \frac{\pa \ }{\pa x^k} (\s^T)^{\ i}_{j} (x,t) \] \, \s^{kj} (x,t) \ := \ b^i (x,t) \ + \ \rho^i (x,t) \ .   \eeq Note that for $\s$ constant we have $\rho = 0$, and hence $b^i= f^i$. See e.g. \cite{Eva,Fre,IW,Kam,Oks} and in particular \cite{Stroock} for details on the Ito-Stratonovich correspondence.

It is natural to wonder if the equivalence between \eqref{eq:Ito}
and \eqref{eq:strato} extends somehow to their symmetries. It
turns out this question is not so simply answered in general, but
the answer is simple in the case of simple symmetries.

First of all, we note that in the case of the Stratonovich
equation \eqref{eq:strato}, its symmetries of the form
\eqref{eq:X}, i.e. its simple deterministic symmetries, are
characterized as solutions to the determining equations \cite{GS}
\begin{eqnarray}  \pa_t \vphi^i \ + \ b^j \, (\pa_j \vphi^i ) \ - \ \vphi^j \, ( \pa_j b^i ) &=& 0 \ , \label{eq:deteqstrat1} \\
\s^j_{\ k} \, (\pa_j \vphi^i ) \ - \ \vphi^j \, (\pa_j \s^i_{\ k} ) &=& 0 \ . \label{eq:deteqstrat2} \end{eqnarray}
For the equation equivalent to
\eqref{eq:Ito}, i.e. taking into account \eqref{eq:itostrat} and with $\rho$ defined in there,
these read \begin{eqnarray}   \pa_t \vphi^i \ + \ f^j \, (\pa_j \vphi^i ) \ - \ \vphi^j \, ( \pa_j f^i ) &=& \rho^j \, (\pa_j \vphi^i ) \ - \ \vphi^j \, ( \pa_j \rho^i ) \ , \label{eq:deteqSI1} \\
\s^j_{\ k} \, (\pa_j \vphi^i ) \ - \ \vphi^j \, (\pa_j \s^i_{\ k} ) &=& 0 \ .  \label{eq:deteqSI2} \end{eqnarray}

\medskip\noindent
{\bf Proposition 1 (Unal \cite{Unal}).} {\it The simple
deterministic symmetries of the Ito equation \eqref{eq:Ito} and
those of the equivalent Stratonovich equation \eqref{eq:strato} --
i.e. if  \eqref{eq:itostrat} holds -- do coincide.}

\medskip\noindent
{\bf Remark 3.} We stress that Unal studied more general cases as
well (so Unal's theorem is more general than the version given
here); as these are not of interest for our present discussion, we
only report the part of his result of direct relevance to us.

On the other hand, we stress that in the same work Unal showed that -- even in the deterministic framework -- the result does \emph{not} extend to non-simple symmetries; in particular if one considers Lie-point symmetries with generator $X = \tau (\pa / \pa t)  + \vphi^i (\pa / \pa x^i) $, the determining equations for the Ito and the associated Stratonovich equation are equivalent if and only if $\tau$ satisfies an additional condition expressed by a third order PDE, more precisely (in our present notation)
$$ \s^k_{\ p} \, \s^{ip} \ \left[ \pa_k \left( \pa_t \tau \ + \ f^j \, (\pa_j \tau ) \ + \ \frac12 \, \s^m_{\ q} \, \s^j_{\ q} \,(\pa_m \pa_j \tau) \right) \right] \ = \ 0 \ ; $$ note this is identically satisfied for $\tau = \tau (t)$  (i.e. for ``acceptable'' cases according to the discussion in \cite{GS}). \EOR
\bigskip

As for simple random symmetries of an Ito equation and of the equivalent Stratonovich one, it was claimed in \cite{GS} that these are not necessarily the same. This statement followed from a (trivially) wrong definition of the Ito Laplacian (see Remark 2 above).

Actually, we are able to prove that Unal's theorem extends to random symmetries.

\medskip\noindent
{\bf Proposition 2.} {\it The simple symmetries of the Ito equation \eqref{eq:Ito} and those of the equivalent Stratonovich equation \eqref{eq:strato} – i.e. if \eqref{eq:itostrat} holds – do coincide.}

\medskip\noindent
{\bf Proof.} The proof follows from direct (and rather boring) computations, reported in the Appendix; they deal with simple random symmetries, but deterministic ones are a special case of the latter, characterized by independence of the functions $\vphi^i (x,t;w)$ on the $w^k$ variables. \EOP

\section{Symmetry of Stratonovich equations and change of coordinates}
\label{sec:redstrat}


As recalled above, see Section \ref{sec:symmsde}, in the case of
deterministic differential equations, equations are identified
with suitable submanifolds -- the solution manifold -- in Jet
spaces (or manifolds); symmetries are vector fields whose
prolongation is tangent to the solution manifold \cite{AVL,CG,Olv1,Olv2,Ste}.
As this is a geometric relation, it is evident that it is
unaffected by changes of coordinates, and we are immediately
guaranteed that a vector field which is computed to be a symmetry
in a given set of coordinates is still such in different
coordinates.

In the case of Ito SDEs, we cannot identify the equation with a
similar solution manifold, and symmetries are usually seen as maps
on the underlying space which leave formally invariant the SDE
itself. As such, i.e. being identified by an \emph{algebraic}
rather than \emph{geometric} relation (see again Section
\ref{sec:symmsde}), we are \emph{not} guaranteed that they are
still symmetry when we pass to a different set of coordinates.
This would endanger the feasibility of a symmetry approach, or at
least force us to heavy computations to check that symmetry are
conserved after each change of coordinates.

A possible way out is to parallel an alternative description of
symmetries for deterministic differential equations, and in
particular for (systems of) first order ODEs\footnote{The same
holds also for higher order equations and systems, of course, but
in the case of first order ones we do not need to introduce and
discuss the \emph{contact structure} \cite{AVL,CG,Olv1,Olv2,Ste}. Moreover we
only need these to discuss the case of Ito equations.}. In fact,
the equations \beql{eq:R1} \frac{d x^i}{d t} \ = \ f^i (x,t) \eeq
are equivalently described in terms of (the vanishing of) the
differential one-forms \beql{eq:R2} \om^i \ := \ \d x^i \ - \ f^i
(x,t) \, \d t \ . \eeq As well known, the property of $X = \xi^i
\pa_i$ of being a symmetry for the system of ODEs (according to
the standard notion of Lie symmetry) is equivalent to the property
that \beql{eq:R3} \[ \L_X (\om^i) \]_{\om = 0} \ = \ 0 \ , \eeq or
equivalently that there are functions $\b^i_j (x,t)$ such that
\beql{eq:R4} \L_X (\om^i) \ = \ \b^i_j (x,t) \ \om^j \ . \eeq
Needless to say, these are also geometric relations and hence
independent of any choice of coordinate system (actually
\eqref{eq:R3} is formulated with no reference to a coordinate
system).

It is thus convenient to see symmetries of SDEs in a similar way;
however, the quantities of interest will transform covariantly
(i.e. in the ``usual'' way, following the chain rule) only if we
consider SDEs in Stratonovich form.

In order to do this, we will see a Stratonovich equation as a
(formal) one-form in the augmented space $(x,t,w)$ -- where $x$
and $w$ can be vector of different dimensions. In this way the
property of being a symmetry will surely be preserved under any
change of coordinates.

\medskip\noindent
{\bf Definition.} {\it The differential equation \eqref{eq:strato}
is represented by the one form \beql{eq:omega} \om^i \ = \ \d x^i \
- \ b^i (x,t) \, \d t \ - \ \s^i_{\ k} (x,t) \circ \d w^k \ .
\eeq}

\medskip\noindent
{\bf Lemma 1.} {\it A vector field \eqref{eq:X} lies in the
annihilator to the form $\om$ representing the equation
\eqref{eq:strato} if and only if it is a symmetry generator for
\eqref{eq:strato}.}

\medskip\noindent
{\bf Proof.} Using Cartan formula \cite{CCL,Nak,NS} for the Lie
derivative of a differential form and \eqref{eq:X}, \eqref{eq:omega}, we have
\begin{eqnarray*}
\L_X (\om^i ) &=& \d (X \interno \om^i ) \ + \ X \interno \d \om^i   \\
&=& \d \( \vphi^i \) \ + \ (\vphi^m \pa_m) \interno (\d t \w (\pa_j b^i) \d x^j \ + \ \d w^k \w \( (\pa_j \s^i_{\ k}) \d x^j \ + \ (\pa_t \s^i_{\ k} \d t \)  \\
&=& (\pa_t \vphi^i) \, \d t \ + \ (\pa_j \vphi^i) \, \d x^j \ - \
\vphi^j \, (\pa_j b^i) \, \d t
\ - \ \vphi^j \,(\pa_j \s^i_{\ k} ) \circ \d w^k \\
&=& (\pa_j \vphi^i) \, \d x^j \ + \ \[ (\pa_t \vphi^i) \ - \ \vphi^j \,
(\pa_j b^i) \] \, \d t \ - \ \vphi^j \,(\pa_j \s^i_{\ k} ) \circ \d
w^k \ . \end{eqnarray*} Restricting now to the annihilator of
$\om^i$, i.e. substituting for $\d x^i$ according to the
Stratonovich equation itself, we get
\begin{eqnarray*}
\[ \L_X (\om^i ) \]_{\om = 0} &=& (\pa_j \vphi^i) \, \[ b^j \, \d t \ + \ \s^j_{\ k} \circ \d w^k \] \ + \ \[ (\pa_t \vphi^i) \ - \ \vphi^j \, (\pa_j b^i) \] \, \d t \ - \
 \vphi^j \,(\pa_j \s^i_{\ k} ) \circ \d w^k \\
&=& \[ (\pa_t \vphi^i) \ + \ b^j \, (\pa_j \vphi^i) \ - \ \vphi^j \,
(\pa_j b^i) \] \, \d t \ + \ \[  \s^j_{\ k} \, (\pa_j
\vphi^i) \ - \ \vphi^j \,(\pa_j \s^i_{\ k} ) \] \circ \d w^k \ .
\end{eqnarray*} We should then require the vanishing of this
expression, which is equivalent to the vanishing of both
expressions in square brackets (the coefficients of, respectively,
$\d t$ and each of the $\d w^k$). These are just the $n + m \cdot
n$ determining equations \eqref{eq:deteqstrat1},
\eqref{eq:deteqstrat2} given above, and obtained in the literature
in the ``standard'' way, i.e. with no reference to differential
forms and Lie derivatives. \EOP

\medskip\noindent
{\bf Lemma 2.} {\it Symmetries of a Stratonovich equation are
preserved under simple changes of coordinates \eqref{eq:smap}.}

\medskip\noindent
{\bf Proof.} This follows at once from having set the Stratonovich
equation in geometrical terms as $\d \om = 0$ and by the
independence of the condition $\L_X (\om) = 0 $ on any coordinate
representation. \EOP

\medskip\noindent
{\bf Lemma 3.} {\it Simple symmetries of an Ito
equation are preserved under simple changes of coordinates \eqref{eq:smap}.}

\medskip\noindent
{\bf Proof.} This follows at once from Lemma 2 and Proposition 1.
\EOP

\medskip\noindent
{\bf Remark 4.} First of all we stress that Lemma 3 refers to all kind (deterministic or random) of simple symmetries; thus it corrects wrong statements in previous work \cite{GGPR,GS} (see again Remark 2).

Our discussion allowed to grant preservation of symmetries (under simple changes of coordinates) for \emph{Stratonovich} equations; it extends to Ito equations only when we are sure of the correspondence between symmetries of an Ito equation and of the equivalent Stratonovich one, i.e. for \emph{simple} symmetries, see Lemma 3.

This also means that when this correspondence is not granted, which is the case for non-simple symmetries (with coefficient $\tau$ depending on the spatial variables – and/or on the Wiener processes in the random case – see Remark 3), it is possible (but not certain) that symmetries of an Ito equation are not preserved under a change of variables. It should however be recalled that maps in which $t$ changes depending on the $x$ and/or the $w$ variables are not acceptable physically; see e.g. the discussion in \cite{GGPR,GS}. \EOR

\section{Application: Kozlov theory}

As mentioned in the Introduction, some of the more substantial results of symmetry theory in the context of SDEs were obtained by R. Kozlov in a series of papers \cite{Koz1,Koz2,Koz3} (see also \cite{GGPR} for a brief summary) in which he investigated the possibility to parallel the classical symmetry treatment of ODEs in the case of SDEs; in particular he found sufficient symmetry conditions guaranteeing the integrability of a scalar SDE, or reducibility of a system of SDEs (or integrability in the case the symmetry algebra is sufficiently large and with the suitable algebraic structure -- exactly as in the ODE case).

Here we will only discuss the problem of integration of a scalar Ito SDE
\beql{eq:koz1} d x \ = \ f (x,t) \, d t \ + \ \s (x,t) \, d w \eeq
which admits a \emph{simple} Lie-point symmetry with generator
\beql{eq:Xkoz} X \ = \ \xi (x,t) \ \pa_x \ ; \eeq
the case of systems would go through similar considerations for what concerns our point in this paper (see also the companion paper \cite{GL2}).

Kozlov's approach is based on finding a change of coordinates, characterized by the coefficient $\xi (x,t)$ of the symmetry vector field, mapping the equation to a manifestly integrable one,
\beql{eq:koz2} d y \ = \ \^f (t) \, d t \ + \ \^\s (t) \, d w \ . \eeq

More specifically, he gives the following result \cite{Koz1}.\footnote{Albeit this is a published result, we provide here a proof of it in order to stress how this depends on an assumption which merits discussion, see Remark 5 below.}

\medskip\noindent
{\bf Proposition 3.} {\it If the SDE \eqref{eq:koz1} admits a vector field \eqref{eq:Xkoz} as a (generator of a) Lie-point symmetry, then it can be transformed into \eqref{eq:koz2}, and hence explicitly integrated.

The integrating change is provided by $y  =  F (x,t)$, where $F$ is the inverse to $\Phi$, i.e. $\Phi [ F (x,t),t] = x$, $F [ \Phi
(y,t), t] =  y$, and \beql{eq:covrel2} \Phi (y,t) \ = \
\int \frac{1}{\vphi (y,t) } \ d y \ . \eeq}

\bigskip\noindent
{\bf Proof.}
The proof of this results amount to a direct computation \cite{Koz1}. In particular, one considers the suggested change of variables; we know it will change the equation \eqref{eq:koz1} into a new Ito equation, which we write as \beql{eq:koz2gen} d y \ = \ \^f (y,t) \, d t \ + \ \^\s (y,t) \, d w \ . \eeq

On the other hand, the vector field $X$ will be written, in the new coordinates, simply as
\beql{eq:Xkozy} X \ = \ \pa_y \ . \eeq
Now, \emph{assuming} this is still a symmetry for the transformed equation it is a simple matter to check from \eqref{eq:deteqIto1}, \eqref{eq:deteqIto2} that necessarily $\^f$ and $\^\s$ do not depend on $y$, i.e. the transformed equation is of the form \eqref{eq:koz2}. \EOP

\bigskip\noindent
{\bf Remark 5.} As stressed above, in the course of the proof one is \emph{assuming} that $X$ is a symmetry of the transformed equation, and actually all of Kozlov theory \cite{Koz1, Koz2, Koz3} is based on such an assumption. As we have remarked in this paper, conservation of symmetries of an Ito equation under changes of coordinates is however not granted {\it apriori}, given that Ito equations are not geometrical objects and transform in their own way.

On the other hand, here we are considering a \emph{(deterministic) simple symmetry} of the equation. Our Lemma 3 guarantees this is indeed preserved under changes of coordinates, and the assumption made above is perfectly justified -- and hence Kozlov theory is well rooted and correct. \EOR

\medskip\noindent
{\bf Example 1.} The Ito equation
\beql{eq:example3} d y \ = \ \( e^{- y} \ - \ \frac12 \, e^{-2 y} \) \, d t \ + \ e^{- y} \, d w \eeq admits the vector field
$ X  =  e^{- y} \pa_y $ as a (Lie-point) symmetry generator. In this case $$  \int \frac{1}{\phi (y)} \, d y \ = \ e^y \ + \ c \ ; $$ hence we should consider the change of variables $x = e^y$. By a straightforward computation, we have indeed that with this variable the initial equation \eqref{eq:example3} reads
\beq d x \ = \ d t \ + \ d w \ ;  \eeq in these variables, $X = \pa_x$. \EOR

\medskip\noindent
{\bf Example 2.} The equation
\beql{eq:example4} d y \ = \ \frac{e^{-t} \, (1 + y^2)^2}{8 y^3} \, \( - 4 y^2 \, + \, e^t (3 y^4 +2 y^2 - 1) \) \, d t \ - \ \frac{(1+y^2)^2}{2 y} \, d w \eeq looks too involved to be studied. However, it admits the vector field
$$ X \ = \ - \, \( \frac{(1+y^2)^2}{2 y} \) \ \pa_y \ = \ \vphi (y) \, \pa_y $$ as a symmetry. In this case we have
$$ \int \frac{1}{\vphi (y)} \ d y \ = \ \frac{1}{1 + y^2} \ . $$
Passing  to the variable
$ x =1/(1+ y^2)$, indeed, the equation \eqref{eq:example4} just reads
\beq d x \ = \ e^{- t} \, d t \ + \ d w \ ; \eeq in these variable we have $X = \pa_x$. \EOR

\section{Conclusions and perspectives}

We noticed that symmetries of an Ito equation are defined only \emph{algebraically}, not \emph{geometrically}; it follows that, contrary to the case of deterministic equations, in this context it is not granted that symmetry are preserved under a change of variables. On the other hand, familiar properties are present in the case of \emph{Stratonovich} stochastic differential equations. We have thus considered the relation between symmetries of an Ito equation and those of the associated Stratonovich equation. In the case of \emph{simple symmetries} the two sets coincide (Proposition 2), and hence preservation of symmetries of the Stratonovich equation entails preservation of symmetries of the corresponding Ito one (Lemma 3).

In particular, this sets on a solid ground the theory developed by Kozlov \cite{Koz1,Koz2,Koz3}, which relates symmetry and integrability --- or reducibility -- of Ito SDEs pretty much as for deterministic ODEs, except that only \emph{simple} symmetries are now relevant.

Note that one could consider more general sets of maps and hence symmetries; in particular, \emph{simple random symmetries} \cite{GS}. In this case the identity between symmetries of an Ito equation and those of the associated Stratonovich equation is also granted by Lemma 3. The extension of Kozlov theory using random symmetries appears therefore possible; it will be discussed in a companion paper \cite{GL2}.

\section*{Acknowledgements}

We are most grateful to an unknown Referee who pointed out the missing term in the formula of the Ito Laplacian \eqref{eq:Delta}, thus avoiding incorrect conclusions. This work follows from the M.Sc. Thesis of CL at Universit\`a di Milano. We thank Francesco Spadaro (EPFL Lausanne) for useful discussions, and an anonymous Referee for constructive criticism. The first version of the paper was written while GG visited SMRI; his work is also supported by GNFM-INdAM.

\begin{appendix}

\section{Proof of Proposition 2}

In this Appendix we give a detailed proof of Proposition 2. We will actually prove a slightly stronger result, which immediately implies Proposition 2.

\medskip\noindent
{\bf Lemma A.1.} {\it The equations \eqref{eq:deteqIto1} and \eqref{eq:deteqSI1} are just the same when restricted to the set of functions satisfying the (identical) equations \eqref{eq:deteqIto2}, \eqref{eq:deteqSI2}.}

\medskip\noindent
{\bf Proof.} Recalling \eqref{eq:itostrat}, and defining
\beq \Sigma (u) \ = \ 2 \ \left[ \vphi^i \, (\pa_j \rho^i) \ - \ \rho^j \, (\pa_j \vphi^i ) \right] \ , \eeq
we can rewrite \eqref{eq:deteqSI1} as
\beq \vphi^i_t \ + \ f^j \, \pa_j \vphi^i \ - \ \vphi^j \, \pa_j f^i \ = \ (1/2) \  \Sigma [ \vphi^i ] \ . \eeq
Thus, comparing  this and \eqref{eq:deteqIto1}, we have to show that
$ \Delta ( \vphi^i ) = \Sigma (\vphi^i )$.
More precisely, we have to show that this holds when we restrict to the set $\mathcal{S}$ of functions $\vphi^i (x,t)$ satisfying \eqref{eq:deteqIto2}, which is just the same as \eqref{eq:deteqSI2} and which we rewrite here for ease of reference:
\beql{eq:DR0} \^\pa_k \vphi^i \ = \ \vphi^p \, (\pa_p \s^i_{\ k} ) \ - \ \s^p_{\ k} \, (\pa_p \vphi^i ) \ . \eeq
Thus, we have to show that
\beql{eq:lemmaA} \vphi \in \mathcal{S} \ \Rightarrow \ \ \Delta ( \vphi^i ) \ = \ \Sigma (\vphi^i )  \ . \eeq

It is immediate -- recalling also that the $\s^i_{\ k}$ do not depend on the $\w^k$ variables -- to obtain some differential consequences of the \eqref{eq:DR0} (which of course only hold on $\mathcal{S}$), i.e.
\begin{eqnarray*}
\pa_j \^\pa_k \vphi^i &=& (\pa_j \vphi^p) \, (\pa_p \s^i_{\ k} ) \ + \ \vphi^p \, (\pa_j \pa_p \s^i_{\ k} \ - \ (\pa_j \s^p_{\ k}) \, (\pa_p \vphi^i) \ - \ \s^p_{\ k} \, (\pa_j \pa_p \vphi^i ) \ ; \\
\^\pa_m \^\pa_k \vphi^i &=& (\^\pa_m \vphi^p) \,(\pa_p \s^i_{\ k}) \ - \ \s^p_{\ k} \, (\pa_p \^\pa_m \vphi^i) \\
&=& \vphi^s \, (\pa_s \s^p_{\ m} ) \, (\pa_p \s^i_{\ k}) \ - \ \s^s_{\ m} \, (\pa_s \vphi^p) \, (\pa_p \s^i_{\ k})
\ - \  \s^p_{\ k} \, (\pa_p \vphi^s) \, (\pa_s \s^i_{\ m} ) \ - \  \s^p_{\ k} \,  \vphi^s \, (\pa_s \pa_p \s^i_{\ m} )  \\
& & \ + \  \s^p_{\ k} \, (\pa_p \s^s_{\ m}) \, (\pa_s \vphi^i) \ + \ \s^p_{\ k} \, \s^s_{\ m} \, (\pa_s \pa_p \vphi^i) \ ; \\
\de^{km} \^\pa_m \^\pa_k \vphi^i &=& \vphi^s \, (\pa_s \s^{pk} ) \, (\pa_p \s^i_{\ k}) \ - \ \s^{sk} \, (\pa_s \vphi^p) \, (\pa_p \s^i_{\ k})
\ - \  \s^p_{\ k} \, (\pa_p \vphi^s) \, (\pa_s \s^{im} ) \ - \  \s^p_{\ k} \,  \vphi^s \, (\pa_s \pa_p \s^{im} )  \\
& & \ + \  \s^p_{\ k} \, (\pa_p \s^{sm}) \, (\pa_s \vphi^i) \ + \ \s^p_{\ k} \, \s^{sm} \, (\pa_s \pa_p \vphi^i) \ . \end{eqnarray*}

With these expressions, and some boring algebra, we can easily compute
\begin{eqnarray*}
\left[ \Delta \vphi^i \right]_{\mathcal{S}} &=& \delta^{km} \, (\^\pa_m \^\pa_k \vphi^i) \ + \ \delta^{mk} \, \s^j_{\ k} \, \s^\ell_{\ m} (\pa_\ell \pa_j \vphi^i ) \ + \ 2 \, \s^j_{\ k} \, (\pa_j \^\pa_k \vphi^i ) \\
&=& \vphi^s \, (\pa_s \s^{pk} ) \, (\pa_p \s^i_{\ k} ) \ - \ \s^{sk} \, (\pa_s \vphi^p) \, (\pa_p \s^i_{\ k}) \ - \ \s^{pk} \, (\pa_p \vphi^s) \, (\pa_s \s^i_{\ k} ) \ - \ \s^{pk} \, \vphi^s \, (\pa_s \pa_p \s^i_{\ k}) \\
& & \ + \ \s^{pk} \, (\pa_p \s^s_{\ k}) \,(\pa_s \vphi^i) \ + \ \s^{kp} \, \s^s_{\ k} \, (\pa_s \pa_p \vphi^i ) \ + \ \s^\ell_{\ k} \, \s^{jk} \, (\pa_\ell \pa_j \vphi^i )  \ + \ 2 \, \s^{jk} \, (\pa_j \vphi^p ) \, (\pa_p \s^i_{\ k}) \\
& & \ + \ 2 \, \s^{jk} \, \vphi^p \, (\pa_p \pa_j \s^i_{\ k}) \ - \ 2 \, \s^{jk} \, (\pa_j \s^p_{\ k}) \, (\pa_p \vphi^i) \ - \ 2 \, \s^{jk} \, \s^p_{\ k} \, (\pa_j \pa_p \vphi^i ) \\
&=& \vphi^s \, \s^{pk} \, (\pa_s \pa_p \s^i_{\ k})  \ + \ \vphi^s \, (\pa_s \s^{pk} ) \, (\pa_p \s^i_{\ k}) \ - \ \s^{sk} \, (\pa_s \s^p_{\ k}) \, (\pa_p \vphi^i ) \ . \end{eqnarray*}

Let us now come to compute $\Sigma (\vphi^i)$. Recalling that
$ \rho^i = (1/2) \left[ (\pa_k \s_j^{\ i})  \s^{kj} \right]$, we get (again with some boring algebra)
\begin{eqnarray*}
\Sigma (\vphi^i ) &=& \vphi^j \, \pa_j [ (\pa_k \s^{ip}) \, \s^k_{\ p} ] \ - \ (\pa_k \s^{pj}) \, \s^k_{\ j} \,(\pa_p \vphi^i ) \\
&=& \vphi^j \, (\pa_j \pa_k \s^{ip} ) \, \s^k_{\ p} \ + \ \vphi^j \, (\pa_k \s^{ip} ) \, (\pa_j \s^k_{\ p} ) \ - \ (\pa_k \s^{pj} ) \, \s^k_{\ j} \, (\pa_p \vphi^i ) \\
&=& \vphi^j \, \s^{kp} \, (\pa_j \pa_k \s^i_{\ p} )  \ + \ \vphi^j \, (\pa_j \s^k_{\ p} ) \, (\pa_k \s^{ip} )  \ - \ \s^{kj} \, (\pa_k \s^p_{\ j} ) \, (\pa_p \vphi^i ) \\
&=& \vphi^s \, \s^{pk} \, (\pa_s \pa_p \s^i_{\ k} )  \ + \ \vphi^s \, (\pa_s \s^{pk} ) \, (\pa_p \s^i_{\ k} )  \ - \ \s^{sk} \, (\pa_s \s^p_{\ k} ) \, (\pa_p \vphi^i ) \ .
\end{eqnarray*}
In the last line we have just renamed dummy indices to emphasize the relation with the result obtained above; note that here we have not used \eqref{eq:DR0}, i.e. this computation holds on any (smooth) function. However, we are interested in applying this general formula to functions lying in $\mathcal{S}$.

We obtain, by direct inspection, that
$$ \left[ \Delta (\vphi^i ) \ - \ \Sigma (\vphi^i ) \right]_{\mathcal{S}}  \ = \ 0 \ . $$ This shows that \eqref{eq:lemmaA} holds and hence completes the proof. \EOP

\end{appendix}

\end{document}